\documentclass[aps,preprintnumbers,nofootinbib,superscriptaddress,11pt]{revtex4}
\usepackage[hypertex]{hyperref}
\usepackage[dvipdfmx]{graphicx} 
\usepackage{amsmath,amssymb,amsfonts,bm,cancel} 

    \def\d{\delta} \def\D{\Delta}       \def\l{\lambda}  \def\m{\mu} \def\n{\nu}        \def\t{\tau}

\newcommand{\Row}[3]{ \begin{pmatrix} #1 & #2 & #3 \end{pmatrix} }
\newcommand{\Column}[3]{ \begin{pmatrix} #1 \\ #2 \\ #3 \end{pmatrix} }

\newcommand{\Diag}[3]{ \begin{pmatrix} #1 & 0 & 0 \\ 0 & #2 & 0 \\ 0 & 0 & #3 \\\end{pmatrix}}

\usepackage{color}

\begin{document}


\title{\large \bf A general formula by $LDL^{T}$ decomposition for the type-I seesaw mechanism}

\preprint{STUPP-21-252}
\author{Masaki J. S. Yang}
\email{yang@krishna.th.phy.saitama-u.ac.jp}
\affiliation{Department of Physics, Saitama University, 
Shimo-okubo, Sakura-ku, Saitama, 338-8570, Japan}


\begin{abstract} 

By performing an approximate spectral decomposition to the inverse mass matrix of the right-handed neutrinos $M^{-1}$, we obtain a basis-independent formula for the type-I seesaw mechanism. 
Mathematically, it is based on the generalized Cholesky (or $LDL^{T}$) decomposition of the symmetric matrix $M^{-1} = L D L^{T}$, with a diagonal matrix $D$ and a lower unitriangular matrix $L$.
Since the diagonalization of $L$ can be inverted 
without solving cubic equations, 
the formula will be useful to investigate general properties of the mechanism, such as  flavor symmetries, generalized $CP$ symmetries, and fine-tunings.  

\end{abstract} 

\maketitle

The type-I seesaw mechanism \cite{Minkowski:1977sc,GellMann:1980v,Yanagida:1979as, Mohapatra:1979ia}, which has a potential to explain neutrino oscillation \cite{Super-Kamiokande:1998kpq} and baryon asymmetry of the universe \cite{Fukugita:1986hr}, is one of the most studied subjects in particle phenomenology. 
However, the general nature of the mechanism is difficult to analyze, because it involves eighteen terms in each matrix element of the mass of light neutrinos $m_{\n}$.
E.~Witten has argued that “the considerations have always been qualitative, and, despite some interesting attempts, there has never been a convincing quantitative model of the neutrino masses” \cite{Witten:2000dt, Xing:2020ald}. 
In order to reduce the number of unnecessary parameters, several parameterizations have been proposed for the seesaw mechanism \cite{Casas:2001sr, Barger:2003gt}. 
In this letter, we derive a concise formula of the type-I seesaw mechanism for Lagrangian parameters in a general basis, by using a spectral decomposition in the first-order perturbation to the inverse mass matrix of the right-handed neutrinos $M_{R}^{-1}$.
%

\vspace{12pt}

First of all, the symmetric mass matrix $M$ of the right-handed neutrinos $\n_{Ri}$ is defined as
\begin{align}
M =
\begin{pmatrix}
 M_{11} & M_{12} & M_{13} \\
 M_{12} & M_{22} & M_{23} \\
 M_{13} & M_{23} & M_{33} \\
\end{pmatrix} \, . 
\label{MR}
\end{align}
These matrix elements $M_{ij}$ are in general complex. 
The inverse matrix of $M$ is
\begin{align}
M^{-1} & = {1\over |M|}
\begin{pmatrix}
 M_{22} M_{33}-M_{23}^2 & M_{13} M_{23}-M_{12} M_{33} & M_{12} M_{23}-M_{13} M_{22} \\
 M_{13} M_{23}-M_{12} M_{33} & M_{11} M_{33}-M_{13}^2 & M_{12} M_{13}-M_{11} M_{23} \\
 M_{12} M_{23}-M_{13} M_{22} & M_{12} M_{13}-M_{11} M_{23} & M_{11} M_{22}-M_{12}^2 \\
\end{pmatrix} \, \label{Minv}  \\
& = 
{1 \over |M|} 
\Column
{(\bm M_{2} \times \bm M_{3})^{T}}
{(\bm M_{3} \times \bm M_{1})^{T}}
{(\bm M_{1} \times \bm M_{2})^{T}} \, , 
\label{Minv2} 
\end{align}
where $|M| = \det M$ and $(\bm M_{i})_{j} \equiv (M_{ij})$ are three-dimensional vectors composed from the rows of the mass matrix.
If the Yukawa matrix of neutrinos $Y$ is hierarchical like the other fermions, $M$ should have a strong hierarchy about the order of $Y\, Y^{T}$. 
Then if we consider
\begin{align}
|M_{33}| \gg |M_{23}| \, , \, |M_{22}|  \gg |M_{13}| \, , \, |M_{12}|\, , \, |M_{11}| \, ,
\label{hier}
\end{align}
the complex mass eigenvalues $M_{i}$ of Eq.~(\ref{MR}) is approximately calculated as 
\begin{align}
( M_{3} \, , \,  M_{2} \, , \, M_{1}) \simeq (M_{33} \, , \, M_{22} - {M_{23}^{2} \over M_{33}} \, , \, {|M| \over M_{2} M_{3}}) \, . 
\label{eigenM}
\end{align}
However, the final result does not formally depend on this approximation. 

Next, an approximate spectral decomposition is performed to Eq.~(\ref{Minv}). 
A matrix $M^{(1)}$ whose rank is one is defined as;
\begin{align}
M^{(1)} \equiv 
\begin{pmatrix}
(M^{-1})_{11} & (M^{-1})_{12} & (M^{-1})_{13} \\[6pt]
(M^{-1})_{12} & \dfrac{(M^{-1})_{12}^{2}}{(M^{-1})_{11}} & \dfrac{(M^{-1})_{12} (M^{-1})_{13}}{(M^{-1})_{11}} \\[12pt]
(M^{-1})_{13} & \dfrac{(M^{-1})_{12} (M^{-1})_{13}}{(M^{-1})_{11}}  & \dfrac{(M^{-1})_{13}^{2}}{(M^{-1})_{11}} \\
\end{pmatrix} \, , 
\end{align}
where $(M^{-1})_{ij}$ is an element of Eq.~(\ref{Minv}).
Explicitly, matrix elements of $M^{(1)}$ are found to be
\begin{align}
{1\over |M|}
\begin{pmatrix}
M_{22} M_{33}-M_{23}^2 & M_{13} M_{23}-M_{12} M_{33} & M_{12} M_{23}-M_{13} M_{22} \\
 M_{13} M_{23}-M_{12} M_{33} & \frac{\left(M_{13} M_{23}-M_{12} M_{33}\right){}^2}{M_{22} M_{33}-M_{23}^2} & \frac{\left(M_{12} M_{23}-M_{13} M_{22}\right) \left(M_{13} M_{23}-M_{12} M_{33}\right)}{M_{22} M_{33}-M_{23}^2} \\
 M_{12} M_{23}-M_{13} M_{22} & \frac{\left(M_{12} M_{23}-M_{13} M_{22}\right) \left(M_{13} M_{23}-M_{12} M_{33}\right)}{M_{22} M_{33}-M_{23}^2} & \frac{\left(M_{12} M_{23}-M_{13} M_{22}\right){}^2}{M_{22} M_{33}-M_{23}^2} \\
\end{pmatrix} \, . 
\end{align}
By using Eq.~(\ref{Minv2}), $M^{(1)}$ can also be written as follows 
\begin{align}
M^{(1)} = {1 \over (M_{22} M_{33} - M_{23}^{2}) |M|}
{(\bm M_{2} \times \bm M_{3})} \otimes
 {(\bm M_{2} \times \bm M_{3})^{T}} \, . 
 \label{M12}
\end{align}
Since $M^{(1)}_{11} = (M^{-1})_{11} = (M_{22} M_{33} - M_{23}^{2}) / |M| \simeq 1 /M_{1}$ holds, this $M^{(1)}$ can be regarded as an approximated matrix of $V_{1} \otimes V_{1}^{T} / M_{1}$ composed of an eigenvector $V_{1}$ that satisfies the relation  $M V_{i}^{*} = V_{i} M_{i}$.  
%

Similarly, we can continue the spectral decomposition for $M_{2}$ and $M_{3}$. 
By subtracting $M^{(1)}$ from $M^{-1}$, 
\begin{align}
\tilde M \equiv 
M^{-1} - M^{(1)} = 
\begin{pmatrix}
 0 & 0 & 0 \\[6pt]
 0 & (M^{-1})_{22}- \dfrac{(M^{-1})_{12}^{2}}{(M^{-1})_{11}} & (M^{-1})_{23} - \dfrac{(M^{-1})_{12} (M^{-1})_{13}}{(M^{-1})_{11}}  \\[12pt]
 0 & (M^{-1})_{23} - \dfrac{(M^{-1})_{12} (M^{-1})_{13}}{(M^{-1})_{11}}  & (M^{-1})_{33} - \dfrac{(M^{-1})_{13}^{2}}{(M^{-1})_{11}} \\
\end{pmatrix} \, . 
\label{7}
\end{align}
A matrix $M^{(2)}$ whose rank is one is extracted from Eq.~(\ref{7}) as
\begin{align}
M^{(2)} \equiv 
\begin{pmatrix}
0 & 0 & 0 \\
0 & \tilde M_{22} & \tilde M_{23} \\
0 & \tilde M_{23} & \frac{\tilde M_{23}^{2}}{\tilde M_{22}}
\end{pmatrix}
= {1\over M_{22} M_{33}-M_{23}^2}
\begin{pmatrix}
 0 & 0 & 0 \\
 0 & {M_{33}} & {- M_{23}} \\
 0 & {-M_{23}} & \frac{M_{23}^2}{M_{33}} \\
\end{pmatrix} \, . 
\end{align}
Since the non-zero eigenvalue of this matrix is
\begin{align}
\frac{M_{23}^2+M_{33}^2}{M_{33} (M_{22} M_{33}-M_{23}^2 )} 
\simeq {1 \over (M_{22} - {M_{23}^2 \over M_{33}}) } \simeq {1\over M_{2}} \, , 
\label{M2}
\end{align}
this $M^{(2)}$ can be regarded as an approximated matrix of  $V_{2} \otimes V_{2}^{T} / M_{2}$ composed of the eigenvector $V_{2}$.
The last remaining element is
\begin{align}
M^{(3)} \equiv  M^{-1} - M^{(1)} - M^{(2)} = \Diag{0}{0}{{1 /  M_{33}}} \,  .
\label{M3}
\end{align}
Then this is almost a matrix made from an eigenvector $V_{3}$.
Note that this spectral decomposition is an approximate one of the first-order perturbation for off-diagonal components $M_{ij}$ with $i \neq j$. The matrix $M^{(i)}$ does not strictly satisfy the condition $M^{(i)} M^{(j)} \propto \d^{ij} M^{(j)}$. 
This decomposition can also be regarded as a reconstruction of $M$ by a perturbative rotation from the diagonalized basis. In this construction, the three vectors are not strictly orthogonal; 
\begin{align}
V_{1}' \propto \Row{(M^{-1})_{11}}{(M^{-1})_{12}}{(M^{-1})_{13}}, ~~
V_{2}' \propto \Row{0}{M_{33}}{- M_{23}}, ~~ 
V_{3}' = \Row{0}{0}{1} \, . 
\end{align}
However, the relation $ M^{(1)} + M^{(2)} + M^{(3)} = M^{-1}$ is exact and no approximation is used.

The Yukawa matrix of neutrinos $Y$ is defined by
\begin{align}
Y =
\begin{pmatrix}
A_1 & B_1 & C_1 \\
A_2 & B_2 & C_2 \\
A_3 & B_3 & C_3 \\
\end{pmatrix} 
\equiv 
\Column{~~~~~ \, \bm Y_{1}^{T} ~~~~~ \, }
{  \bm Y_{2}^{T} }
{  \bm Y_{3}^{T} } \, , 
\end{align}
where $(\bm Y_{i})_{j} \equiv Y_{ij} = (A_{i}, \,  B_{i} , \, C_{i})$ are 3-dimensional vectors with 
complex parameters $A_{i}, B_{i}, C_{i}, (i = 1,2,3)$. 
By neglecting the vacuum expectation value of the Higgs field $v$, 
the mass dimension of $Y$ becomes one. 

By considering the seesaw mechanism with this decomposition 
(Eqs.~(\ref{M12}), (\ref{M2}) and (\ref{M3})\,), 
the mass matrix $m$ for the light neutrinos $\n_{i}$ is found to be
\begin{align}
m &= Y (M^{(1)} + M^{(2)} + M^{(3)}) Y^{T} 
\equiv m^{(1)} + m^{(2)} + m^{(3)}\\
&= {|M^{(23)}| \over \det M }
\begin{pmatrix}
 {a_1^2} & {a_1 a_2} & {a_1 a_3} \\
 {a_1 a_2} & {a_2^2} & {a_2 a_3} \\
 {a_1 a_3} & {a_2 a_3} & {a_3^2} \\
\end{pmatrix}
+ 
{M_{33} \over |M^{(23)}|} 
\begin{pmatrix}
b_{1}^{2} & b_{1} b_{2} & b_{1} b_{3} \\
b_{1} b_{2} & b_{2}^{2} & b_{2} b_{3} \\
b_{1} b_{3} & b_{2} b_{3} & b_{3}^{2}
\end{pmatrix}
+ 
{1\over M_{33}}
\begin{pmatrix}
 {C_1^2} & {C_1 C_2} & {C_1 C_3} \\
 {C_1 C_2} & {C_2^2} & {C_2 C_3} \\
 {C_1 C_3} & {C_2 C_3} & {C_3^2} \\
\end{pmatrix} \, , \label{formula}
\end{align}
where $|M^{(23)}| \equiv M_{22} M_{33}-M_{23}^2$ and 
\begin{align}
a_{i} \equiv {\det ( \bm Y_{i} , \, \bm M_{2} , \, \bm M_{3}) \over |M^{(23)}|} , ~~
b_{i} \equiv {(\bm Y_{i} \times \bm M_{3})_{1} \over M_{33}}
= Y_{i2} - Y_{i3} {M_{23} \over M_{33}} \, . 
\label{cross}
\end{align}
This is an exact expression in a general basis without approximations. 
Since $a_{i}$ is proportional to the scalar triple product, it represents a component of $\bm Y_{i}$ in a direction orthogonal to $\bm M_{2}$ and $\bm M_{3}$ (or paralell to $\bm M_{2} \times \bm M_{3}$).
From the hierarchy~(\ref{hier}) or (\ref{eigenM}), $a_{i}$ and $b_{i}$ are approximately equal to the first and second elements of $\bm Y_{i}$. 
Substituting $(a_{i} , \, b_{i} ) \sim (A_{i} , \, B_{i} )$, we obtain a similar representation as 
Ref.~\cite{Barger:2003gt}.
\begin{align}
m_{ij} & \simeq {1\over M_{1}} A_{i} A_{j} + {1\over M_{2}} B_{i} B_{j} + {1\over M_{3}} C_{i} C_{j} \, . 
\end{align}
The equality holds for diagonal $M$. 
Each matrix roughly represents contributions from $M_{1,2,3}\,$.

Mathematically, it corresponds to the generalized Cholesky (or $LDL^{T}$) decomposition of the symmetric matrix $M$; 
\begin{align}
\tilde M^{-1} \equiv L M^{-1} L^{T} 
= \Diag{|M^{(23)}| \over \det M}{M_{33} \over |M^{(23)}|}{1\over M_{33}} \, , ~~~ 
L = 
\begin{pmatrix}
 1 & 0 & 0 \\
 \frac{M_{13} M_{23}-M_{12} M_{33}}{M_{23}^2-M_{22} M_{33}} & 1 & 0 \\
 \frac{M_{13}}{M_{33}} & \frac{M_{23}}{M_{33}} & 1 \\
\end{pmatrix} \, .
\end{align}
Here, $L$ is a lower unitriangular matrix such that all diagonal elements are one. 
From here, if we define $\tilde Y \equiv Y L^{-1} = (\bm a \, , \bm b \, , \bm C)$ with $(\bm a \, , \bm b \, , \bm C)_{i} \equiv (a_{i} \, , b_{i} \, , C_{i}) \, $, 
the following holds obviously; 
\begin{align}
m = Y M^{-1} Y^{T} = \tilde Y \tilde M^{-1} \tilde Y^{T} \, .
\end{align}

With the normalized eigenvectors $v_{i}$ satisfying $m \, v_{i} = m_{i} \, v_{i}^{*}$ (that is equivalent to $U^{T} m U = m^{\rm diag}$ with a unitary matrix $U$), the mass eigenvalues can be expressed as
\begin{align}
m_{i} = 
{|M^{(23)}| \over \det M }
(\bm a \cdot v_{i})^{2}
+ 
{M_{33} \over |M^{(23)}|} 
(\bm b \cdot v_{i})^{2}
+ 
{1\over M_{33}}
(\bm C \cdot v_{i})^{2} \, , 
\end{align}
where $(\bm a)_{i} \equiv a_{i}, \, (\bm b)_{i} \equiv b_{i}, \, (\bm C)_{i} \equiv C_{i}$ and
$(\bm u \cdot \bm v) \equiv u_{1} v_{1} + u_{2} v_{2} + u_{3} v_{3}$ is the inner product without Hermitian conjugation.

The general formula (\ref{formula}) has implications for several phenomenologies.

{\it 1. flavor structure. }
In the third term $m^{(3)}$ of Eq.~(\ref{formula}), 
at most only the 33 element contributes to $m$ 
because the hierarchy of $M$~(\ref{hier}) and a relation $|C_{3}| \gg |C_{2}| \gg |C_{1}|$ that holds for many Yukawa matrices. 
Therefore, the first and second terms $m^{(1,2)}$ make important contributions in the seesaw mechanism.
On the other hand,  the trimaximal mixing of the MNS matrices \cite{Harrison:2002kp,Harrison:2004he,Friedberg:2006it,Lam:2006wy,Bjorken:2005rm, He:2006qd, Grimus:2008tt, Gautam:2016qyw} is observed with good accuracy.
If there is no nontrivial relation between $a_{i}$ and $b_{i}$, 
the trimaximal mixing requires $a_{1} \simeq a_{2} \simeq a_{3}$ or $a_{1} + a_{2} + a_{3} \simeq 0$ ($b_{i}$ must satisfy another condition of $a_{i}$ to avoid $m_{2} \simeq 0$ or $m_{1,3} \simeq 0$). 
In particular, by setting $A_{1} \simeq A_{2} \simeq A_{3} \simeq B_{1} \simeq \d, ~ B_{2} \simeq - B_{3} \simeq \l $ with $|\l| \gtrsim |\d|$, the matrices $Y$ and $m$ become 
\begin{align}
Y \simeq 
\begin{pmatrix}
\d & \d & C_{1} \\
\d & \l & C_{2} \\
\d & - \l & C_{3}
\end{pmatrix} \, , 
~~~ 
m \simeq {M_{33} \over |M^{(23)}|} 
\begin{pmatrix}
\d^{2} + \tilde \d^{2} & \d \l + \tilde\d^{2}  & - \d \l + \tilde\d^{2} \\
 \d \l + \tilde\d^{2}&  \l^{2} + \tilde\d^{2} & - \l^{2} + \tilde\d^{2} \\
- \d \l + \tilde\d^{2} &  -\l^{2} + \tilde\d^{2}  & \l^{2} + \tilde\d^{2}
\end{pmatrix} \, ,
\label{18}
\end{align}
where $\tilde \d^{2} \simeq  {|M^{(23)}|^{2} \, \d^{2} / M_{33} \, |M|}  \sim  {M_{2} \, \d^{2}/ M_{1}}$ has an absolute value of about $m_{2}$.
The form of $Y$ recalls the cascade hierarchy $C_{1} \simeq \d, \, C_{2} \simeq - \l $ \cite{Dorsner:2001sg, Haba:2008dp} suggested by the CKM matrix and is rather desirable from a viewpoint of unified models. 
%
%
Similar results have been obtained by (constrained) sequential dominance \cite{King:1998jw,King:1999cm,King:1999mb,King:2002nf,Antusch:2004gf, King:2005bj} and previous studies of fine-tuning in the seesaw mechanism \cite{Dermisek:2004tx,Sayre:2007ps, Meloni:2012sx}. 

{\it 2. naturalness and fine-tunings. } 
Since the chiral and lepton number symmetries are restored in the limit of $Y$ and $M$ being zero, the quantum corrections are naturally small if these parameters are small\footnote{Even if some of these parameters are large, Yukawa couplings of leptons without strong interactions are hardly renormalized. Also, even if $Y_{33}$ is as large as the top Yukawa coupling, $Y_{33} \sim m_{t}$, the flavor-dependent contribution is $O(0.1)$, about 10\% \cite{Xing:2020ijf}.}~\cite{tHooft:1979bh}. 
Nevertheless, since the type-I seesaw mechanism involves many subtractions, 
there is still a possibility of fine-tuning, in which large parameters cancel each other out. 
The naturalness of the seesaw mechanism in this meaning has been studied \cite{Dermisek:2004tx,Sayre:2007ps, Meloni:2012sx}. 

To avoid such a large cancellation in Eq.~(\ref{formula}), 
parameters $a_{i}$, $b_{i}$, and $C_{i}$ to stay within certain ranges.
By defining the mass scale of $m$ by $\sqrt{|\D m^{2}_{3l}|}$, such conditions are given by 
\begin{align}
\left| {\det ( \bm Y_{i} , \, \bm M_{2} , \, \bm M_{3}) \over \sqrt { |M^{(23)}| \det M} } \right | \, , \, 
\left|{(\bm Y_{i} \times \bm M_{3})_{1} \over \sqrt{M_{33} (\bm M_{2} \times \bm M_{3})_{1}}} \right | \, , \, 
\left| {C_{i} \over \sqrt{M_{33}}} \right |  \lesssim \sqrt[4]{|\D m^{2}_{3l}|} \, .
\label{natural}
\end{align}
Since $M_{1}$ is the lightest eigenvalue, the restriction is strongest for $a_{i}$. 
In other words, the component of $\bm Y_{i}$ in the direction of $\bm M_{2} \times \bm M_{3}$,  which is approximately equal to $Y_{i1} = A_i$, must be the same order of magnitude for any $i$.
However, even if some components of $\bm Y_{i}$ are large, Eq.~(\ref{natural}) still valid if the alignments $\bm Y_{i} \propto \bm M_{2,3}$ holds approximately \cite{Yang:2021byq}.  The lopsided texture \cite{Sato:1997hv} is included in this case. 
Since there are several equivalent ways to perform the spectral decomposition and $LDL^{T}$ decomposition, a more systematic analysis is needed to identify such conditions. 

{\it 3. $CP$ symmetry. }
In the case of the normal hierarchy (NH), either the first or second term in Eq.~(\ref{formula}) can be a dominant matrix with rank one that produces $m_{3}$.
If $m^{(2)}$ is such a dominant matrix, the neutrino mass $m$ has approximate $CP$ symmetry in a basis where phases of $b_{i}$ and $M_{33} / |M^{(23)}|$ are removed by redefinition of fields. 
Even if the second eigenvalue $m_{2}$ is not neglected,  
for example, imposition of the $\m - \t$ symmetry \cite{Fukuyama:1997ky,Lam:2001fb,Ma:2001mr,Balaji:2001ex,Koide:2002cj} and NH yields a solution; 
\begin{align}
b_{1} = 0 \, , ~~ a_{3} = a_{2} \, , ~~ b_{3} = - b_{2} \, . 
\end{align}
The form of $m$ becomes
\begin{align}
m \simeq 
{|M^{(23)}| \over \det M }
\begin{pmatrix}
 {a_1^2} & {a_1 a_2} & {a_1 a_2} \\
 {a_1 a_2} & {a_2^2} & {a_2^{2}} \\
 {a_1 a_2} & {a_2^{2}} & {a_2^2} \\
\end{pmatrix}
+ 
{M_{33} \over |M^{(23)}|} 
\begin{pmatrix}
0 & 0 & 0 \\
0 & b_{2}^{2} & - b_{2}^{2} \\
0 & - b_{2}^{2} & b_{2}^{2}
\end{pmatrix} \, . 
\end{align}
In this case, the mass eigenvalues $m_{\n i}$ (of the complex values in general) are
\begin{align}
(m_{\n 1} \, , m_{\n 2}\, , m_{\n 3} ) = 
(0 \, , {|M^{(23)}| \over \det M} (a_{1}^{2} + 2 \,a_{2}^{2}) \, , {M_{33} \over |M^{(23)}|} 2 \, b_{2}^{2} ) \, .  
\label{eigenmn}
\end{align}
In Eq.~(\ref{eigenmn}), 
there exists a generalized $CP$ symmetry (GCP) \cite{Ecker:1983hz, Gronau:1985sp, Feruglio:2012cw, Holthausen:2012dk} such as the diagonal reflection symmetry 
$R \, m_{\n}^{*} \, R = m_{\n}$ with $R = {\rm diag} (-1,1,1)$ 
\cite{Yang:2020qsa,Yang:2020goc,Yang:2021smh,Yang:2021xob} 
 related to the phases of eigenvalues $m_{\n i}$. 
Similar considerations can be made for the inverted hierarchy, and some characteristic solutions can involve a generalized $CP$ symmetry.
Furthermore, by applying the general formula to the minimal seesaw model \cite{Ma:1998zg, King:1998jw, Frampton:2002qc}, we can derive the general condition for $m$ to have $CP$ symmetry \cite{Yang:2022wch}. 
Such general conditions can be investigated for other flavor symmetries and seem to be useful.

To conclude,  
by performing an approximate spectral decomposition to the inverse mass matrix of the right-handed neutrinos $M^{-1}$, we obtain a concise formula for the type-I seesaw mechanism in a general basis. 
Mathematically, it is based on the generalized Cholesky (or $LDL^{T}$) decomposition of the symmetric matrix $M^{-1} = L D L^{T}$, with a diagonal matrix $D$ and a lower unitriangular matrix $L$. 
Since the diagonalization of $L$ can be inverted without solving cubic equations, 
the formula will be useful to investigate general properties of the mechanism, such as   flavor symmetries, generalized $CP$ symmetries, and fine-tunings.  
Similar expressions are expected for other analogs of the seesaw mechanism.

{\it Acknowledgement:}
This study is financially supported 
by JSPS Grants-in-Aid for Scientific Research
No.~18H01210, No. 20K14459,  
and MEXT KAKENHI Grant No.~18H05543.


\end{document}